# Affine Extension of Galilean Conformal Algebra in 2+1 Dimensions

Ali Hosseiny and Shahin Rouhani

Physics Department,

Sharif University of Technology,

Tehran PO Box 11165-9161, Iran.

Email: alihosseiny@physics.sharif.edu , srouhani@sharif.ir

**Abstract**

We show that a class of non-relativistic algebras including non centrally-extended Schrodinger algebra and Galilean Conformal Algebra (GCA) has an affine extension in 2+1 hitherto unknown. This extension arises out of the conformal symmetries of the two dimensional complex plain. We suggest that this affine form may be the symmetry that explains the relaxation of some classical phenomena towards their critical point. This affine algebra admits a central extension and maybe realized in the bulk. The bulk realization suggests that this algebra may be derived by looking at the asymptotic symmetry of an AdS theory. This suggests that AdS/CFT duality may take on a special form in four dimensions.

## 1. Introduction

The AdS/CFT correspondence [1-3] since its inception has been source of many fruitful ideas. In its simplest interpretation the AdS/CFT correspondence conjectures a gravity theory on a d+1 dimensional Anti-de Sitter space-time, be dual to a d dimensional conformally invariant theory on the boundary of the AdS. This conjecture has recently been extended to low-energy physical systems [4,5]. For a recent review of the extension of the AdS/CFT ideas to the condensed matter setting see [6]. The natural question that arises is why a conjecture should hold true in a setting that it was not intended for. Intuitively this translocation stands a good chance of being true since it corresponds to a limit where the velocity of light becomes very large and there is no reason to believe that it should breakdown any where en route as c tends to infinity. However to make sure and check the validity of the conjecture in its new setting, one may check three things. Firstly how does the correspondence fair in the string theory framework [7-11]. Secondly, can you derive the correlation functions from the dual theory correctly, which was done in the early days [4,5] also see a recent paper [12]. Thirdly, do you get the right symmetry groups in the asymptotes of the gravity dual and the boundary theory? This means that these theories need to have non-relativistic symmetries both on the boundary and correspondingly on the asymptotes of the gravitational dual [13,14]. In the simplest case, the non-relativistic symmetry of the boundary is taken to be the Schrodinger symmetry which is the symmetry of the free Schrodinger equation [15-23]. This symmetry algebra admits of two extra generators as well as the usual Galilean operations. The extra symmetries are the *dilation* and *inversion*; hence the nomenclature of conformal is used. In fact the conformal invariance is justified due to the geometric interpretation offered by the Newton-Cartan structure [24-30]. Other symmetries on the boundary are possible, comprising the Galilean group as a subgroup, in fact there is a class of conformal Galilean algebras, generalizing the Schrodinger algebra. They may be obtained by contraction [31] of the relativistic algebra o(d,2). In this way the Galilean Conformal Algebra (GCA) is obtained. The algebra GCA has a long history. For an early reference to GCA and the Schrodinger algebra see [32] later GCA and its affine extension were investigated in [21] and more recently investigated in [24] and [33]. Unlike the Schrodinger algebra, GCA scales time and space in the same manner. The non-relativistic Navier-Stokes equation has been suggested as an example of a physical system which is not relativistic but has GCA symmetry. This is a likely candidate since space and time scale in the same manner, however this issue has been debated. [24-34] .The algebra GCA does not admit a mass term, so it is the massless version of Galilean symmetry. An interesting aspect of GCA is that it can be extended to an affine algebra [24], the affine extension has also been known for the Schrodinger algebra [19,20]. Though GCA does not admit mass central charge, it does admit exotic central charge in $d = 2 + 1$ [39-43]. The existence of an infinite extension signals the existence of gravity dual with asymptotic symmetries, but one has to be careful using the terminology of "AdS" space, these spaces are different. The crucial question is the natural affine extension, and its geometrical interpretation. In this article we argue that in the particular case of $d = 2$, that is when space-time is 2+1 dimensional, an alternative infinite extension of non-relativistic algebras exists. This affine extension corresponds to the natural possibility of allowing the space symmetries develop into the familiar infinite extension (the Virasoro algebra). This paper is organized as follows; in section 2 we consider the various non relativistic algebras which have the Galilean algebra as subalgebra. In section 3 we look at the affine extensions of these algebras. In section 4 we point out an alternative affine extension in 2+1 dimensions. In section 5 we provide the algebra that holds in the bulk.

## 2. Non- relativistic conformal symmetries

Non-relativistic conformal algebras may be obtained by contracting [31] relativistic conformal algebras. For this purpose one or more directions of the space-time are scaled with a parameter which then is allowed to tend to zero, [24,33-34]. Physically this is equivalent to letting the speed of light becoming very large. For this process we set:

$$x \to \varepsilon x, \qquad t \to t \ , \qquad while \ \ \varepsilon \to 0 \ . \qquad (2.1)$$

This results in non-relativistic algebras which contains the Galilean algebra:

$$P_i = \partial_i, \qquad H = -\partial_t, \qquad B_i = t\partial_i, \qquad J_{ij} = -(x_i \partial_j - x_j \partial_i) \ , \qquad (2.2)$$

as a subalgebra. However the interesting question is what we get extra and on top of the Galilean algebra. The extra bits lead to the Schrodinger and conformal Galilean algebras.

## 2.1 Schrodinger Group

Let us begin with the Schrodinger group, the group of symmetries of Schrodinger's equation [15-23]. This symmetry algebra is larger than the Galilean symmetry. The free Schrodinger equation;

$$\left(i\partial_t + \frac{1}{2m}\partial_i\partial_i\right)\psi = 0, \tag{2.3}$$

is invariant under the set of transformations:

$$\vec{r} \to \vec{r}\,' = \frac{\mathcal{R}\vec{r}+\vec{V}t+\vec{a}}{\gamma t + \delta}, \qquad t \to t' = \frac{\alpha t + \beta}{\gamma t + \delta}, \qquad \alpha\delta - \beta\gamma = 1 \;. \tag{2.4}$$

In addition to Galilean algebra, we have now two more transformations: dilation and inversion (the corresponding transformation of special conformal transformation).

$$D = -(2t\partial_t + x_i\partial_i), \tag{2.5}$$

$$K = -(tx_i\partial_i + t^2\partial_t). \tag{2.6}$$

In Schrodinger conformal transformation, the dilation operator; $D$; scales space and time in an anisotropic manner:

$$x_i \to \lambda\, x_i, \qquad\qquad t \to \lambda^2\, t. \tag{2.7}$$

This is in contradistinction to the usual conformal transformations, where space and time are scaled in a similar manner. The inversion operator; $K$; transforms space and time as follows:

$$x_i \to \frac{x_i}{(1+\mu t)}, \qquad\qquad t \to \frac{t}{(1+\mu t)}. \tag{2.8}$$

It is playing the role of special conformal transformations but time component remains unmodified.

The non-vanishing commutators of the corresponding algebra are

$$[H, B_i] = -P_i, [K, P_i] = B_i, \quad [H, K] = D, \quad [D, B_i] = -B_i,$$

$$[D, P_i] = -P_i, [D, K] = -2K, \qquad\qquad [J_{ij}, J_{kl}] = SO(d),$$

$$[J_{ij}, B_k] = -(B_i\delta_{jk} - B_j\delta_{ik})\,, \qquad\qquad [J_{ij}, P_k] = -(P_i\delta_{jk} - P_j\delta_{ik})\,. \tag{2.9}$$

The algebra admits a central extension, which is the mass:

$$[B_i, P_j] = m\delta_{ij}\,, \tag{2.10}$$

It is therefore clear that the Galilean algebra is a subalgebra of the Schrodinger algebra, also that Schrodinger algebra is a subalgebra of the relativistic conformal algebra in one higher dimension .Therefore as the symmetry of the boundary it can be realized via a gravity dual.

## 2.2 Galilean conformal algebra (GCA)

An alternate contraction which does not respect centre of mass frame and does not accept mass central charge, while still keeps Galilean algebra as a subgroup is Galilean conformal algebra. To gain this algebra, we first look at the relativistic conformal algebra:

$$P_i = \partial_i, \qquad P_0 = -\partial_0 = H, \qquad J_{ij} = -(x_i \partial_j - x_j \partial_i),$$
$$J_{0i} = t\partial_i - x_i \partial_t, \qquad D = -(x_\mu \partial^\mu),$$
$$K_\mu = -(2x_\mu (x.\partial) - (x.x)\partial_\mu). \tag{2.11}$$

After contraction (2.1) we are left with some new operators:
$$B_i = J_{0i} = t\partial_i, \qquad K = K_0 = -(2tx_i \partial_i + t^2 \partial_t), \qquad K_i = t^2 \partial_i. \tag{2.12}$$

The rest of the operators keep their relativistic infinitesimal form. This includes the dilation operator $D$. So, unlike the Schrodinger algebra, in Galilean conformal algebra, dilation scales time and space equally. The algebra $(J_{ij}, H, D, K, P_i, K_i, B_i)$ has Galilean algebra $(J_{ij}, H, P_i, B_i)$ as subalgebra. The full algebra is

$$[K, B_i] = K_i, \quad [K, P_i] = 2B_i, \quad [K, H] = 2D, \quad [K, D] = K, \quad [D, K_i] = -K_i,$$
$$[D, P_i] = P_i, \quad [D, H] = H, \quad [H, K_i] = -2B_i, \qquad [H, B_i] = -P_i,$$
$$[J_{ij}, K_l] = -(K_i \delta_{jl} - K_j \delta_{il}) \quad [J_{ij}, B_l] = -(B_i \delta_{jl} - B_j \delta_{il}),$$
$$[J_{ij}, P_l] = -(P_i \delta_{jl} - P_j \delta_{il}). \tag{2.13}$$

The other commutators vanish. This algebra does not admit a mass central charge. However in $d = 2 + 1$, it admits a different central charge [35]:

$$[B_i, B_j] = \theta \epsilon_{ij}, \qquad [P_i, K_j] = -2\theta \epsilon_{ij}. \tag{2.14}$$

Physical interpretation of this charge; called "exotic"; has been of some interest, whereas some authors have kept mass central charge along with exotic charge [36-43].

## 2.3 Spin-*l* Galilean class

The last two versions of Galilean algebras were obtained via two different manners. First one was the Schrodinger equation's symmetry and the other one was obtained by contracting conformal algebra. However yet there is another way to obtain a conformal Galilean symmetry. In this approach we look at the symmetries of Galilean $d + 1$ dimension for $d > 2$. Here we have a *time dimension* plus a *d-dimensional Euclidean space*. First we look at symmetries of these dimensions.

Space is Euclidean and can carry a conformal symmetry $SO(d+1,1)$. The corresponding algebra would be:

$$P_i = \partial_i, \qquad D = -x_i\partial^i, \qquad J_{ij} = -(x_i\partial_j - x_j\partial_i),$$

$$K_i = -(2x_i(x.\partial) - (x.x)\partial_i). \tag{2.15}$$

Time is one dimensional and can carry a $SL(2,R)$ which is the generator of global transformations:

$$t \to t' = \frac{\alpha t + \beta}{\gamma t + \delta}, \tag{2.16}$$

However since it is one dimension it can carry the affine form symmetry of $SL(2,R)$ or Witt algebra generated by:

$$T^n = -t^{n+1}\partial_t. \tag{2.17}$$

Now we would like to add to these algebras. We are left with three constraints. Firstly, in classical physics, space cannot be mixed with time. In other words we can't have any operator as $f(x)\partial_t$. However time for sure can be rotated into space via the Galilean boost:

$$B_i = t\partial_{x_i}. \tag{2.18}$$

So, this Galilean boost stays at the center of all Galilean symmetries. Secondly inspired by the dynamical index, we allow space and time scaling correlated via the dilation operator:

$$D = -lx_i\partial^i - t\partial_t. \tag{2.19}$$

This operator scales time and space as:

$$t \to \lambda t, \qquad \vec{x} \to \lambda^l \vec{x}. \tag{2.20}$$

Finally we seek closure for the algebra. These three constraints result in an "Spin-$l$" algebra presented in [21-23]:

$$H = -\partial_t, \qquad D = -(t\partial_t + lx_i\partial_i), \quad K = -(2ltx_i\partial_i + t^2\partial_t),$$

$$J_{ij} = -(x_i\partial_j - x_j\partial_i), \qquad P_i^n = (-t)^n\partial_i. \tag{2.21}$$

Where $i = 1, ..., d$ and $n = 0, ..., 2l$. In these algebras $l$ can be half integer hence the terminology of spin has been used for it, though there is no correspondence with the familiar spin. The case of $l = \frac{1}{2}$ corresponds to non-centrally extended Schrodinger symmetry and $l = 1$ corresponds to GCA.

These algebras correspond to the following transformations:

$$\vec{r} \to \vec{r}' = \frac{\mathcal{R}\vec{r} + t^{2l}\vec{c}_{2l} + \cdots + t^2\vec{c}_2 + t\vec{c}_1 + \vec{c}_0}{(\gamma t + \delta)^{2l}}, \qquad t \to t' = \frac{\alpha t + \beta}{\gamma t + \delta}. \tag{2.22}$$

where $R \in SO(d)$, $\vec{c}_n \in \mathbb{R}^d$ and $\alpha\delta - \beta\gamma = 1$. Now we are left with a question. What happens for $d = 2 + 1$ where we have 2 Virasoro algebras?

$$L_m = -z^{m+1}\partial_z, \qquad\qquad \bar{L}_m = -\bar{z}^{m+1}\partial_{\bar{z}}. \tag{2.23}$$

Should we still sacrifice two infinite algebraic elements for having a closed algebra? This is the question we would expand on in section 4.

### 3. Affine extensions

We now come to the question of affine extension of the above. The algebras introduced above admit affine extensions; meaning there exist infinite dimensional algebras of which these algebras are closed finite subalgebras. The crucial question is however: what restricts the affine extension? It seems to us that there should exist a geometrical interpretation limiting the form of the affine extension. This constraint is offered in many ways. Either by the underlying space-time, for instance the Newton-Cartan interpretation gives a geometrical setting, or that certain symmetry is inherited from the dual gravity in the shape of asymptotic symmetries and induced on to the boundary. Alternatively the equation concerned as it stands, such as the Schrodinger equation, admits of nonlinear invariance.

### 3.1 Schrodinger Virasoro

Schrodinger algebra admits a Virasoro-like affine extension [19]. Together with mass central charge we have the following infinitesimal transformations:

$$T^n = -t^{n+1}\partial_t - \frac{1}{2}(n+1)t^n x \partial_x - \frac{1}{4}n(n+1)\mathcal{M}t^{n-1}x^2,$$

$$P^m = -t^{m+\frac{1}{2}}\partial_x - (m+\frac{1}{2})t^{m-1/2}x\mathcal{M},$$

$$M^n = -\mathcal{M}t^n . \tag{3.1}$$

Where $n \in \mathbb{Z}$ and $m \in \mathbb{Z} + \frac{1}{2}$. In this format:

$$P^{-\frac{1}{2}} = -\partial_x = -P, \qquad P^{\frac{1}{2}} = B,$$

$$T^{-1} = H, \qquad\qquad T^0 = D/2, \qquad\qquad T^1 = K, \tag{3.2}$$

are the generators of the Schrodinger symmetry mentioned in section 2.1 in 1+1. The generators given above form the Schrodinger-Virasoro algebra:

$$[T^n, T^m] = (n+m)T^{n+m}, \qquad\qquad [T^n, P^m] = (\frac{n}{2} - m)P^{n+m}.$$

$$[T^n, M^m] = -mM^{n+m} , \qquad [P^n, P^m] = (n-m)M^{n+m} ,$$

$$[P^n, M^m] = [M^n, M^m] = 0 . \tag{3.3}$$

### 3.2 GCA affine extension

GCA as well admits an affine extension. First we set

$$T^{-1} = H, \qquad T^0 = D, \qquad T^1 = K,$$

$$M_i^{-1} = P_i, \qquad M_i^0 = B_i, \qquad M_i^1 = K_i. \tag{3.4}$$

Now, we can define the general form

$$T^n = -(n+1)t^n x_i \partial_i - t^{n+1} \partial_t,$$

$$M_i^n = t^{n+1} \partial_i,$$

$$J_{ij}^n = -t^n (x_i \partial_j - x_j \partial_i). \tag{3.5}$$

We observe that the affine extension of GCA for $d = 1+1$ can be obtained via a direct contraction from relativistic conformal algebra in this dimension which is the Virasoro algebra. Let us go to complex coordinates, $z = \frac{x+it}{\sqrt{2}}$, then we have two copies of affine $SL(2,R)$ generated by:

$$L_n = -z^{n+1} \partial_z , \qquad \bar{L}_n = -\bar{z}^{n+1} \partial_{\bar{z}}. \tag{3.6}$$

For contraction we set

$$x \to \frac{x}{c} , \qquad t \to t . \tag{3.7}$$

getting:

$$L_n = -\left(\frac{x/c+it}{\sqrt{2}}\right)^{n+1} \left(\frac{c\partial_x - i\partial_t}{\sqrt{2}}\right)$$

$$= -\left(\frac{1}{\sqrt{2}}\right)^{n+2} [(it)^{n+1} c\partial_x - i(it)^{n+1} \partial_t + (n+1)(it)^n x\partial_x + o\left(\frac{1}{c}\right)] . \tag{3.8}$$

Now we define for even n:

$$M^n = i\left(\frac{\sqrt{2}}{i}\right)^n \left[\frac{L_n - \bar{L}_n}{c}\right] = t^{n+1} \partial_x + o\left(\frac{1}{c}\right) ,$$

$$T^n = (\sqrt{2})^n i^{-n-1} [L_n + \bar{L}_n] = -t^{n+1} \partial_t - (n+1) t^n x \partial_x + o\left(\frac{1}{c}\right). \tag{3.9}$$

And for odd n :

$$M^n = -i(\tfrac{\sqrt{2}}{i})^n \left[\tfrac{L_n + \bar{L}_n}{c}\right] = t^{n+1}\partial_x + o(\tfrac{1}{c}) \quad ,$$

$$T^n = (\sqrt{2})^n i^{-n-1}[L_n - \bar{L}_n] = -t^{n+1}\partial_t - (n+1)t^n x \partial_x + o(\tfrac{1}{c}) \quad . \tag{3.10}$$

So, after letting $c \to \infty$ we directly end up with affine GCA and observe that two $SL(2,R)$ algebras are exactly the full GCA algebra.

The affine algebra in arbitrary dimensions is

$$[T^m, T^n] = (m-n)T^{m+n}, \qquad [T^m, J_a^n] = -nJ_a^{m+n}, \qquad [J_a^m, J_b^n] = f_{abc} J_c^{m+n}$$

$$[T^m, M_i^n] = (m-n)M_i^{m+n}, \qquad [M_i^m, M_j^n] = 0,$$

$$[M_i^m, J_{jk}^n] = (M_j^{m+n}\delta_{ik} - M_k^{m+n}\delta_{ij}). \tag{3.11}$$

Here $f_{abc}$ is $SO(d)$ structure constants. It has been shown that this affine form can be realized in the bulk [24].

## 4. Affine extension of Spin-*l* Galilean symmetry in $d = 2+1$

The affine form presented in the last section arises via the fact that in Galilean models time is decoupled from space. One result of this fact is that we can have this affine form for any dimension $d+1$. However, when looking at the case of $d = 2+1$ we would have another dimension in which we can have an affine form. We think that the necessity of this affine extension comes from the fact that in 2-dimensional critical systems, representations of *spatial* Virasoro play a crucial role. For example 2-dimensional Ising model sits in a $c = \tfrac{1}{2}$ representation of Virasoro algebra and tricritical Ising model sits in a $c = \tfrac{7}{10}$ representation of spatial Virasoro. It is while that Hamiltonian of these 2-dimensional critical systems mainly does respect Galilean symmetry when adding time. So, it seems to us that we need another affine form than presented in the last section to explain relaxation of these physical systems near a critical point. When restricted to $2+1$ dimensions we would like to keep our infinite algebraic symmetry in our space which is produced by two spatial Virasoro or

$$L_n = -z^{n+1}\partial_z, \qquad \bar{L}_n = -\bar{z}^{n+1}\partial_{\bar{z}}. \tag{4.1}$$

Having this request added to Spin-*l* Galilean symmetry, we end up with the algebra generated by the following:

$$L_m^n = -t^n z^{m+1}\partial_z \quad , \qquad \bar{L}_m^n = -t^n \bar{z}^{m+1}\partial_{\bar{z}},$$

$$T^n = -(t^{n+1}\partial_t + l(n+1)t^n(z\partial_z + \bar{z}\partial_{\bar{z}})). \tag{4.2}$$

In this representation the Spin-*l* Galilean algebra is:

$$H = -\partial_t = T^{-1}, \qquad\qquad D = -t\partial_t - l(z\partial_z + \bar{z}\partial_{\bar{z}}) = T^0,$$

$$K = -t^2\partial_t - 2lt(z\partial_z + \bar{z}\partial_{\bar{z}}) = T^1, \; \{P_i^n\} = \{t^n\partial_z, t^n\partial_{\bar{z}}\} = \{L_{-1}^n, \bar{L}_{-1}^n\},$$

$$J = i(z\partial_z - \bar{z}\partial_{\bar{z}}) = -i(L_0^0 - \bar{L}_0^0). \tag{4.3}$$

The commutators are as follows:

$$[L_m^k, L_n^l] = (m-n)L_{m+n}^{k+l}, \qquad [\bar{L}_m^k, \bar{L}_n^l] = (m-n)\bar{L}_{m+n}^{k+l},$$

$$[L_m^k, \bar{L}_n^l] = 0, \qquad [T^m, T^n] = (m-n)T^{m+n},$$

$$[L_m^k, T^n] = (k + mln + ml)L_m^{k+n}, \quad [\bar{L}_m^k, T^n] = (k + mln + ml)\bar{L}_m^{k+n}. \tag{4.4}$$

A trivial central charge can be assigned to this algebra;

$$[T^m, T^n] = (m-n)T^{m+n} + \frac{c}{12}m(m^2-1)\delta_{m+n,0}. \tag{4.5}$$

We have not been able to derive another central charge for this algebra, but we have no reason to claim that other central charges are not admissible. We like to remark that it was the requirement of closure that lead to the $l$ being half integer. However, when working on affine form this closure is no longer natural. This means that the dynamical index $z = 1/l$ may take on more general values.

## 5. Bulk Symmetry

In this section we try to look at the bulk symmetry for our affine form for the special case of $l = 1$ or GCA. Following [24] we look at AdS$_4$, with the hope that the algebra we have found may be extended to some asymptotic algebra, and find that this is indeed the case. It is simple enough to restrict our attention to the metric offered by [24] in four dimensions. The isometries of this metric are:

$$P_i = \partial_i, \qquad B_i = (t-r)\partial_i - x_i\partial_t,$$

$$K_i = (t^2 - 2tr - x_j^2)\partial_i + 2tx_i\partial_t + 2rx_i\partial_r + 2x_ix_j\partial_j,$$

$$J_{ij} = -(x_i\partial_j - x_j\partial_i), \quad H = -\partial_t, \qquad D = -t\partial_t - r\partial_r - x_i\partial_i,$$

$$K = -(t^2 + x_i^2)\partial_t - 2r(t-r)\partial_r - 2(t-r)x_i\partial_i. \tag{5.1}$$

If we now do a contraction to the non-relativistic limit we get:

$$P_i = \partial_i, \qquad B_i = (t-r)\partial_i,$$

$$K_i = (t^2 - 2tr)\partial_i, \qquad J = -(x_1\partial_2 - x_2\partial_1),$$

$$H = -\partial_t, \qquad D = -t\partial_t - r\partial_r - x_i\partial_i,$$

$$K = -t^2\partial_t - 2(t-r)(r\partial_r + x_i\partial_i). \tag{5.2}$$

Since we would like to have an affine extension, we interpret our elements in the language of our $L_m^n$:

$$\{P_i\} = \{L^0_{-1}, \bar{L}^0_{-1}\}, \qquad \{B_i\} = \{(t-r)L^0_{-1}, (t-r)\bar{L}^0_{-1}\},$$

$$\{K_i\} = \{(t^2 - 2tr)L^0_{-1}, (t^2 - 2tr)\bar{L}^0_{-1}\}, \quad J = -i(L^0_0 - \bar{L}^0_0),$$

$$H = -\partial_t \quad , \qquad D = -t\partial_t - r\partial_r + (L^0_0 + \bar{L}^0_0),$$

$$K = -t^2\partial_t - 2(t-r)(r\partial_r - L^0_0 - \bar{L}^0_0) \quad . \tag{5.3}$$

Now we define our space time affine extension

$$T^{(m)} = -t^{m+1}\partial_t - (m+1)(t^m - mrt^{m-1})(z\partial_z + \bar{z}\partial_{\bar{z}} + r\partial_r),$$

$$L^{(n)}_{(m)} = -(t^n - nrt^{n-1})z^{m+1}\partial_z, \qquad \bar{L}^{(n)}_{(m)} = -(t^n - nrt^{n-1})\bar{z}^{m+1}\partial_{\bar{z}}. \tag{5.4}$$

Here z stands for our 2-dimensional space which we do contraction on.

Commutation relations up to the lowest order of $r$ are as follows:

$$\left[L^{(k)}_{(m)}, L^{(l)}_{(n)}\right] = (m-n)L^{(k+l)}_{(m+n)}, \qquad \left[\bar{L}^{(k)}_{(m)}, \bar{L}^{(l)}_{(n)}\right] = (m-n)\bar{L}^{(k+l)}_{(m+n)},$$

$$\left[L^{(k)}_{(m)}, \bar{L}^{(l)}_{(n)}\right] = 0, \qquad \left[T^{(m)}, T^{(n)}\right] = (m-n)T^{(m+n)},$$

$$\left[L^{(k)}_{(m)}, T^{(n)}\right] = (k + mn + m)L^{(k+n)}_{(m)}, \left[\bar{L}^{(k)}_{(m)}, T^{(n)}\right] = (k + mn + m)\bar{L}^{(k+n)}_{(m)} . \tag{5.5}$$

We observe that this algebra, to the lowest order in $r$ is the same algebra as the boundary affine (4.4). However what is missing is a geometric structure, it would be nice to construct a geometrical structure in the spirit of [30], furthermore to show that our affine algebra indeed arises out of the asymptotic Killing vectors.

## 6. Concluding Remarks

We have seen that a class of non-relativistic algebras may have an affine extension in space which has not been yet noticed. It's been suggested [19] that NRCFT can explain dynamics of some non-relativistic critical phenomena near or over critical point. We think that our affine form may explain the dynamics of some non-relativistic ones in 2 dimensions. Those ones that respect Galilean boost and as well sit in representations of spatial Virasoro algebra at their critical point. The key would appear when the representation theory and Verma modules of this algebra are worked out. The fact that this algebra can be extended to the bulk suggests that there may exist a higher dimensional theory which has this symmetry as a dual in $2+1$ dimensions.

**Acknowledgements**

We are indebted to Mohsen Alishahiha, Mohammad Rajabpour and Saman Moghimi for interesting discussions and reading the manuscript. Also we thank Peter Horvathy for comments on an earlier version.

## 7-References